\documentclass[sigconf]{acmart}
\usepackage{subfigure}
\usepackage{graphicx}

\usepackage{balance}
\def\BibTeX{{\rm B\kern-.05em{\sc i\kern-.025em b}\kern-.08emT\kern-.1667em\lower.7ex\hbox{E}\kern-.125emX}}
    
\copyrightyear{2019} 
\acmYear{2019} 
\setcopyright{acmcopyright}
\acmConference[MM '19]{Proceedings of the 27th ACM International Conference on Multimedia}{October 21--25, 2019}{Nice, France}
\acmBooktitle{Proceedings of the 27th ACM International Conference on Multimedia (MM '19), October 21--25, 2019, Nice, France}
\acmPrice{15.00}
\acmDOI{10.1145/3343031.3356049}
\acmISBN{978-1-4503-6889-6/19/10}

\begin{document}

\fancyhead{}
  % do not delete this code.

% The "title" command has an optional parameter, allowing the author to define a "short title" to be used in page headers.
\title{A Hybrid Control Scheme for Adaptive Live Streaming}

% The "author" command and its associated commands are used to define the authors and their affiliations.
% Of note is the shared affiliation of the first two authors, and the "authornote" and "authornotemark" commands
% used to denote shared contribution to the research.
\author{Huan Peng}
\orcid{0000-0001-6648-9353}
\affiliation{%
  \institution{Communication University of China}
  \streetaddress{No. 1 Dingfuzhuang East Street}
  %\city{Chaoyang Qu}
  %\state{Beijing Shi}
  %\country{China}
  \postcode{100024}
}
\email{hpeng@cuc.edu.cn}

\author{Yuan Zhang}
\affiliation{%
  \institution{Communication University of China}
  \streetaddress{No. 1 Dingfuzhuang East Street}
  %\city{Chaoyang Qu}
  %\state{Beijing Shi}
  %\country{China}
}
\email{yzhang@cuc.edu.cn}

\author{Yongbei Yang}
\affiliation{%
  \institution{Communication University of China}
  \streetaddress{No. 1 Dingfuzhuang East Street}
  %\city{Chaoyang Qu}
  %\state{Beijing Shi}
  %\country{China}
}
\email{vrwhatvdo@cuc.edu.cn}

\author{Jinyao Yan}
\affiliation{%
 \institution{Communication University of China}
 \streetaddress{No. 1 Dingfuzhuang East Street}
 %\city{Chaoyang Qu}
 %\state{Beijing Shi}
 %\country{China}
}
\email{jyan@cuc.edu.cn}

%
% By default, the full list of authors will be used in the page headers. Often, this list is too long, and will overlap
% other information printed in the page headers. This command allows the author to define a more concise list
% of authors' names for this purpose.
\renewcommand{\shortauthors}{Huan Peng and Yuan Zhang, et al.}

%
% The abstract is a short summary of the work to be presented in the article.
\begin{abstract}
The live streaming is more challenging than on-demand streaming, because the low latency 
is also a strong requirement in addition to the trade-off between video quality and jitters 
in playback. To balance several inherently conflicting performance metrics and improve the 
overall quality of experience (QoE), many adaptation schemes have been proposed. Bitrate 
adaptation is one of the major solutions for video streaming under time-varying 
network conditions, which works even better combining with 
some latency control methods, such as 
adaptive playback rate control and frame dropping. However, it still 
remains a challenging problem to design an algorithm to combine these adaptation schemes together. 
To tackle this problem, we propose a hybrid control scheme for adaptive live streaming, namely HYSA,  
based on heuristic playback rate control, latency-constrained bitrate control and 
QoE-oriented adaptive frame dropping. The proposed scheme utilizes Kaufman’s Adaptive 
Moving Average (KAMA) to predict segment bitrates for better rate decisions. 
Extensive simulations demonstrate that HYSA outperforms most of the existing adaptation
schemes on overall QoE.
\end{abstract}

  

%
% Keywords. The author(s) should pick words that accurately describe the work being
% presented. Separate the keywords with commas.

%
% A "teaser" image appears between the author and affiliation information and the body 
% of the document, and typically spans the page. 
%   \Description{Enjoying the baseball game from the third-base seats. Ichiro Suzuki preparing to bat.}

%
% This command processes the author and affiliation and title information and builds
% the first part of the formatted document.
\maketitle

\section{Introduction}
Recent years have seen tremendous growth of live streaming applications. 
Different from on-demand streaming, live streaming has tight latency constraints.
It's very challenging to reduce the latency 
while maintaining high video quality and smooth playback. 
Bitrate adaptation is the most common solution for improving the QoE of video streaming 
under time-varying network conditions. 
However, the existing adaptive bitrate algorithms, 
such as BOLA \cite{Bola}, RobustMPC \cite{RobustMPC}, Pensieve \cite{Pensieve} and Oboe \cite{Oboe}, 
haven't taken the latency into account.
In addition to adapting the video bitrate, playback rate control and 
frame dropping are always utilized to reduce the latency in live streaming. 
Mingfu Li et al. \cite{adaptiveplayback} employed playback rate adaptation, 
and Miller et al. \cite{LOLYPOP} 
and Shen Y et al. \cite{framedropping} adopted frame dropping to reduce the latency. However, 
these methods lack the capability to balance between latency and video quality, without 
fully considering all aspects of QoE.

In this paper, we propose HYSA, an effective hybrid control scheme to realize playback rate adaptation, 
bitrate adaptation and frame dropping adaptation. First, the playback rate is adaptively adjusted with a 
buffer-based heuristic method. Then, taking advantage of the 
playback rate decisions and   
KAMA-based predicted segment bitrates,
we propose the latency-constrained bitrate adaptation scheme to make optimal bitrate decisions for  
the QoE-oriented frame dropping adaptation scheme in the next step. 
Our extensive simulation results demonstrate that the proposed HYSA outperforms 
existing adaptation schemes on the overall QoE.

The rest of the paper is organized as follows. Section 2 introduces the system framework 
in live streaming scenario and the simulation platform. Section 3 describes the details of 
the proposed hybrid control scheme. In Section 4, the performance of our method is 
evaluated comprehensively. Section 5 concludes this paper.

\section{System Overview}

\begin{figure}[t]
  \centering  
  \includegraphics[width=\linewidth]{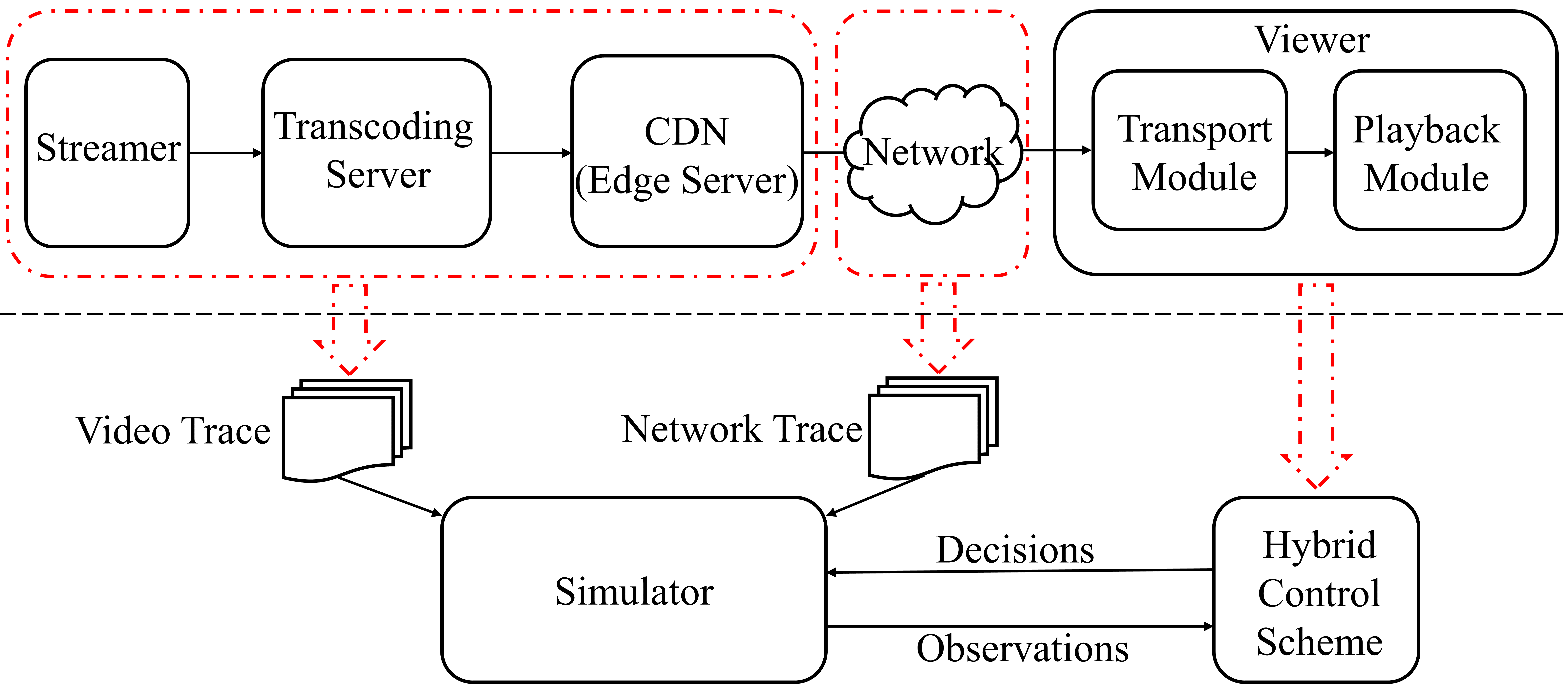}
  \caption{\label{fig:system architecture}Architecture of live streaming system and the simulator}
  \Description{Architecture of live streaming system and simulator.}
\end{figure}

Figure \ref{fig:system architecture} depicts a typical scenario of live streaming. 
The video frames generated in real-time 
are uploaded to a transcoding server, which re-encodes the video into multiple 
representations, each at a different bitrate. These representations are 
then transmitted to CDN (Content Delivery Network) nodes, which act as edge servers. 
The client decides which representation to download from one of the CDN nodes based 
on some state information, such as buffer occupancy and throughput.  
Besides, the client can adaptively adjust its playback rate and skip some frames to reduce the latency.

The simulator simulates the downloading of video frames under various network 
conditions and the adaptive playback of a player. The simulator takes 
video trace, network trace and decisions from the hybrid control scheme as inputs. The video trace
records the size of video frames and their time of arrival at CDN, while the network 
trace simulates throughputs of the downloading network. The simulator collects 
observations after downloading every frame, then the control scheme makes decisions after 
downloading a complete group of pictures (GOP) by taking advantage of these observations. 
Hereinafter we use a segment to refer to a GOP.

\section{Hybrid Control Scheme}
\begin{figure}[t]
  \centering
  \includegraphics[width=\linewidth]{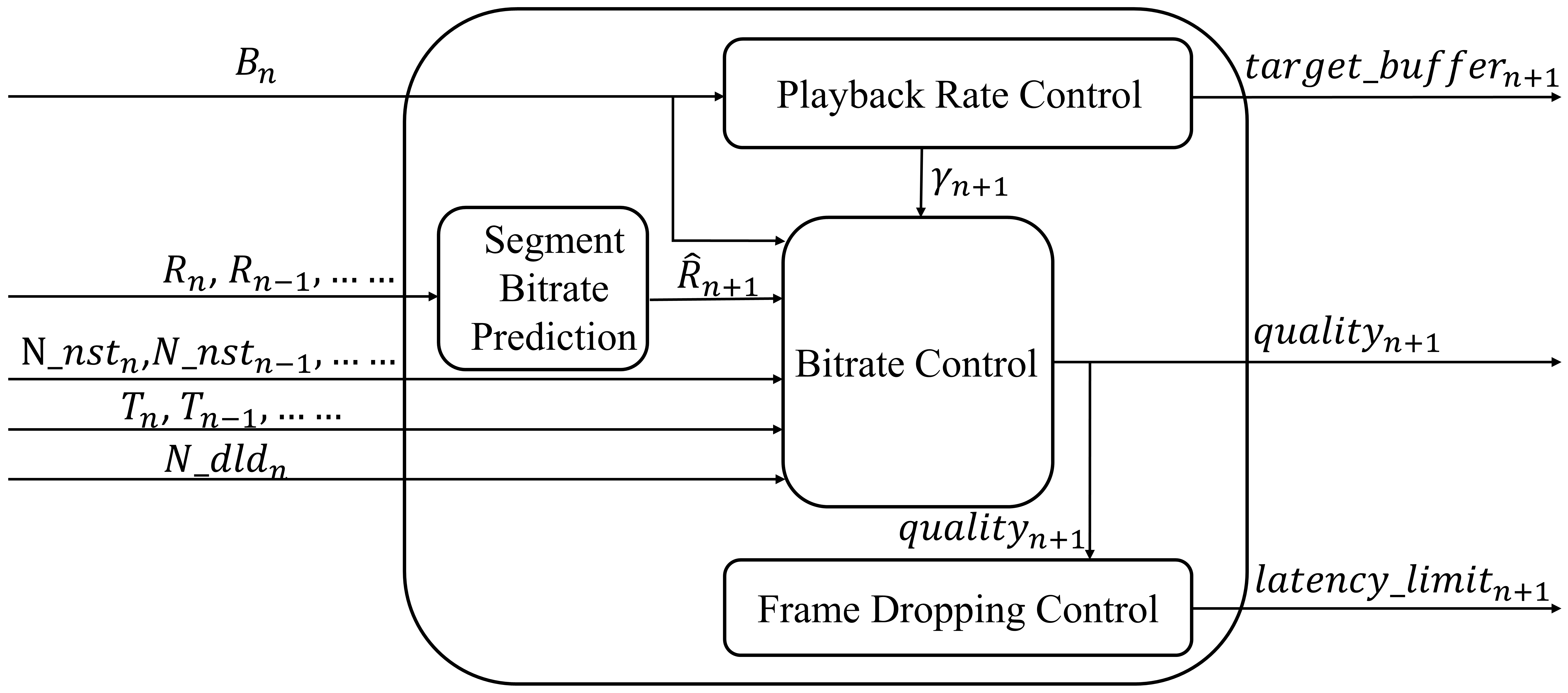}
  \caption{\label{fig:algorithm_architecture}Overview of the proposed hybrid control scheme}
  %A block diagram of the proposed hybrid control scheme for adaptive live streaming}
  \Description{Architecture of the proposed hybrid control scheme for live streaming}
\end{figure}

In this section, we describe the details of the proposed hybrid control scheme -- HYSA, 
which consists of segment bitrate prediction module, playback rate control module, 
bitrate control module and frame dropping control module, 
as shown in Figure \ref{fig:algorithm_architecture}.
% whose architecture is shown in Figure \ref{fig:algorithm_architecture}.
% 3.1 QoE model
\subsection{QoE model}
% The QoE metric is of vital improtance for the adaptive control scheme design. 
Many studies have highlighted the critical role that QoE plays in the design of adaptation schemes. 
Here we refer to the QoE model specified by the grand challenge in 
ACM MM 2019\footnote{https://www.aitrans.online/MMGC/}, 
which mainly focuses on five performance metrics: video quality, rebuffering, 
latency, frame skipping and quality switching. Their impacts on QoE are notated by $QoE_{quality}$, $QoE_{rebuf}$, 
$QoE_{latency}$, $QoE_{skip}$ and $QoE_{switch}$ respectively. 
The overall QoE is calculated as follows:
\begin{gather}
\begin{align}
    QoE = &\   \space QoE_{quality} + QoE_{rebuf} + QoE_{latency} + QoE_{skip} \nonumber \\
                   &+ QoE_{switch} \space \nonumber \\
    % QoE =  \space QoE_{quality}& + QoE_{rebuf} + QoE_{latency} + QoE_{skip} + QoE_{switch} \space &\nonumber \\
        = &\  \sum_{k=1}^{K} (p_{q}V_{k} d_{f} - p_{r} t_{k}^r - p_{l} l_k - p_{s} t_{k}^s- p_{w} | V_{k} - V_{k-1}|)% \nonumber \\
                  %& - p_{w} | V_{k} - V_{k-1}|)
\end{align} 
\end{gather}
where $K$ is the total number of frames. The coding bitrate 
of frame $k$ is notated by $V_k$, while $d_f$ is its 
length. $t_{k}^r$ and $t_{k}^s$ denote the rebuffering duration and video length skipped 
when downloading frame $k$ respectively, and $l_k$ is the latency. $p_q$, $p_r$, 
$p_l$, $p_s$, $p_w$ are weight factors used to describe the importance of corresponding QoE metrics. 
$|V_{k}-V_{k-1}|$ describes the quality variation of two adjacent video frames.
% \begin{equation}
%   \rho_{n} = 
%   \begin{cases}
%     0, & \text{if} \ current\ frame\ is\ I\ frame \\
%     1, & \text{otherwise}
%   \end{cases}
% \end{equation}

\subsection{Segment Bitrate Prediction}
\begin{figure}[t]
  \centering
  \includegraphics[width=\linewidth]{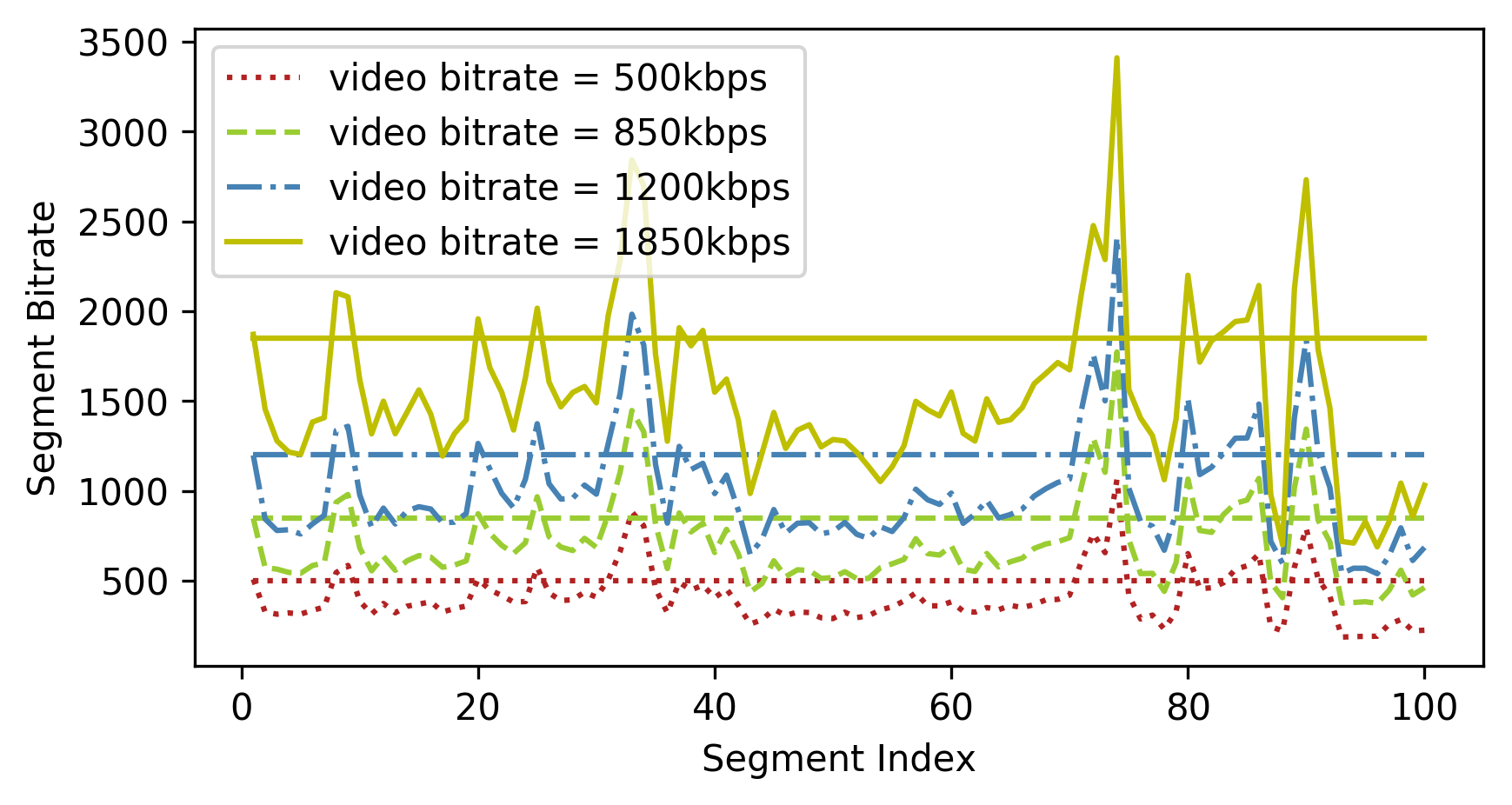}
  \caption{\label{segment_bitrate}Actual bitrates of segments at variable coding bitrates}
  \Description{Real Bitrates of Segments at Variable Coding Bitrates}
\end{figure}

The next download duration can be estimated if the next segment’s actual bitrate is known. 
However, it isn’t realistic to get information about the video that hasn’t been generated in live 
streaming. Therefore, most of the existing algorithms 
estimate the next download duration using the coding bitrate of the upcoming segment 
instead of its actual bitrate, 
ignoring the fact that segment's actual bitrate varies significantly for a given coding bitrate, 
as indicated by Figure \ref{segment_bitrate}. 
Besides, we can observe that for segments in two videos with the same content but different 
qualities, the ratios of their actual bitrates have a similar trend with the ratios 
of their coding bitrates.

% Obviously, using the coding bitrate to estimate the time of next downloading period 
% is extremely inaccurate. 
Assume that each segment is edcoded at $M$ different bitrates, 
and let $V_{n,m}$ and $R_{n,m}$ be the coding bitrate and the actual  
bitrate of segment $n$ at quality level $m$, satisfying $V_{n,m_{1}} < V_{n,m_{2}}$, $R_{n,m_{1}} < R_{n,m_{2}}$, 
$\forall m_{1}<m_{2}$. For the $n$-th segment already downloaded at quality $q_{n}$, 
we can estimate its actual bitrates of other quality levels 
based on the above observation:
\begin{equation}
  R_{n,m} \approx \frac{V_{n,m}}{V_{n,q_{n}}}R_{n,q_{n}}, m\neq q_{n} \ and \ m \in [1,M]
\end{equation}

Here we employ Kaufman’s Adaptive Moving Average (KAMA) to predict the actual bitrates of the 
upcoming segment. The bitrate of the next segment $n+1$ at quality level $m$, denoted as $\hat{R}_{n+1,m}$,  
can be predicted as follows:
\begin{equation}
%   \hat{R}_{n+1,m} = \sum_{i=n-n_1}^{n} \alpha_{i} R_{i,m}
    \hat{R}_{n+1,m} = (1-SC_{n}) \hat{R}_{n,m} + SC_{n} R_{n,m}
\end{equation}
The smoothing factor $SC_{n}$ is dynamically calculated for every sample, i.e. segment bitrate. 
To get the smoothing factor, we first set two boundaries for it, based on the method of calculating 
smoothing factor in Exponential Moving Average (EMA):
\begin{equation}
    SC_{slowest}=\frac{2}{l_{max}+1}
\end{equation}
\begin{equation}
    SC_{fastest}=\frac{2}{l_{min}+1}
\end{equation}
The $l_{max}$ and $l_{min}$ are the number of samples for the slowest and fastest EMA respectively. 
Then we calculate the efficiency ratio $ER_{n}$, which shows the efficiency of sample fluctuations.
% \begin{equation}
%     ER_{n}=\frac{direction_{n}}{volatility_{n}}
% \end{equation}
% \begin{equation}
%     direction_{n}=|R_{n,m} - R_{n-N_{1},m}|
% \end{equation}
% \begin{equation}
%     volatility_{n}=\sum_{i=0}^{N_{1}-1} |R_{n-i,m} - R_{n-i-1,m}|
% \end{equation}
\begin{equation}
  ER_{n}=\frac{|R_{n,m} - R_{n-N_{1},m}|} {\sum_{i=0}^{N_{1}-1} |R_{n-i,m} - R_{n-i-1,m}|}
\end{equation}
where $N_1$ specifies the number of samples used for calculating $ER_n$, and $ER_n$ 
is always between 0 and 1. Using $ER_n$ and two boundaries, the $SC_n$ can be derived as below:
\begin{equation}
    SC_{n}=[ER_{n} (SC_{fastest} - SC_{slowest}) + SC_{slowest}]^2
\end{equation}

\subsection{Playback Rate Control}
%In the playback rate control module, we difine an upper threshold $B_{max}$ 
%and a lower threshold $B_{min}$ to describe a buffer interval that 
%the player can play buffered video at normal playback rate. 
%When the buffer occupancy is below $B_{min}$, the player will slow down the playback rate to 
%0.95, while speeding up the playback rate to 1.05 when the buffer occupancy is above $B_{max}$. 
%Then we use $B_{target}$ to represent the target buffer occupancy to resume playback when interruptions occur. 
%Besides, a variable named $target_buffer$ is introduced to describe a combination of $B_{min}$, $B_{max}$, and $B_{target}$, and it can be 
%marked as $[B_{min}, B_{target}, B_{max}]$.
In the playback rate control module, the target buffer is introduced to help control 
playback rate $\gamma$. The so-called target buffer is marked as $[B_{min}, B_{target}, B_{max}]$, where $B_{min}$ 
and $B_{max}$ form a buffer interval that player can play buffered video at normal playback rate, and ${B_{target}}$ means 
the target buffer occupancy to resume playback when interruptions occur. 
When the buffer occupancy is below $B_{min}$, the player will slow down the playback rate to 
0.95, while speeding up the playback rate to 1.05 when the buffer occupancy is above $B_{max}$. 
In the grand challenge, the target buffer can only be set to 
0 or 1, i.e. $[B_{min}^0, B_{target}^0, B_{max}^0]$ or $[B_{min}^1, B_{target}^1, B_{max}^1]$, 
satisfying $B_{min}^0 < B_{min}^1 < B_{target}^0 < B_{target}^1 < B_{max}^0 < B_{max}^1$.
% as shown in Figure \ref{target_buffer}.

The heuristic playback rate control module decides which target buffer to choose solely depending 
on current buffer occupancy $B_{n}$. Five cases are considered, as described below:

Case 1:$B_n<B_{min}^0$, which means there is a substantial risk of interruptions, 
and $\gamma=0.95$ whichever target buffer is chosen. To restart playback as 
soon as possible, the target buffer is set to 0 due to smaller $B_{target}$.

Case 2:$B_n \in [B_{min}^0,B_{min}^1)$, which means stalls may be encountered 
even though it is less possible than Case 1. Therefore, the target buffer 
is set to 1 to make player slow down the playback rate to 0.95.

Case 3:$B_n \in [B_{min}^1,B_{max}^0)$, which means buffer occupancy 
remains in a reasonable interval, and $\gamma=1.0$ whichever target buffer is chosen.

Case 4:$B_n \in [B_{max}^0,B_{max}^1)$, which means the latency is relatively large 
due to buffered video. Therefore, the target buffer is set to 
0 to speed up the playback rate to 1.05.

Case 5:$B_n \geq B_{max}^1$, which means large latency caused by buffered video,   
and $\gamma$ is equal to 1.05 whichever target buffer is chosen.

Based on the discussions above, the playback rate can 
be adjusted by the target buffer as follows:
\begin{equation}
  target\_buffer_{n+1} = 
  \begin{cases}
    1, & \text{if} \ B_n \in [B_{min}^0,B_{max}^0) \\
    0, & \text{otherwise}
  \end{cases}
\end{equation}
\begin{equation}
  \gamma_{n+1} = 
  \begin{cases}
    0.95, & \text{if} \ B_n \in [0,B_{min}^1) \\
    1.0, & \text{if} \ B_n \in [B_{min}^1,B_{max}^0) \\
    1.05, & \text{otherwise}
  \end{cases}
\end{equation}

% 3.4 bitrate control
\begin{table*}[t]
  %\caption{Avarege QoE metrics that each adaptive scheme achieves on the entire dataset}
  \caption{Performance comparison of the adaptation schemes}
  \label{QoE_metrics}
  \begin{tabular}{ccccccc}
    \toprule
    Method & $QoE_{overall}$ & $QoE_{quality}$ & $QoE_{rebuf}$ & $QoE_{latency}$ & $QoE_{skip}$ & $QoE_{switch}$ \\
    \midrule
    HYSA      & 2424.04 & 3548.00 & -418.92 & -608.32 & -86.28 & -10.44\\
    HYSA-N    & 2336.33 & 3398.36 & -398.11 & -571.27 & -82.28 & -10.37\\
    MPC       & 2000.44 & 3056.23 & -255.98 & -759.17 & -30.23 & -10.41\\
	  DTTB      & 2038.34 & 3472.63 & -410.20 & -952.57 & -63.72 & -7.80\\
    \bottomrule
  \end{tabular}
\end{table*}

\begin{figure*}
  \centering
  \subfigure[Sports Video]{
    \begin{minipage}[b]{0.3\linewidth}
    \includegraphics[width=1\linewidth]{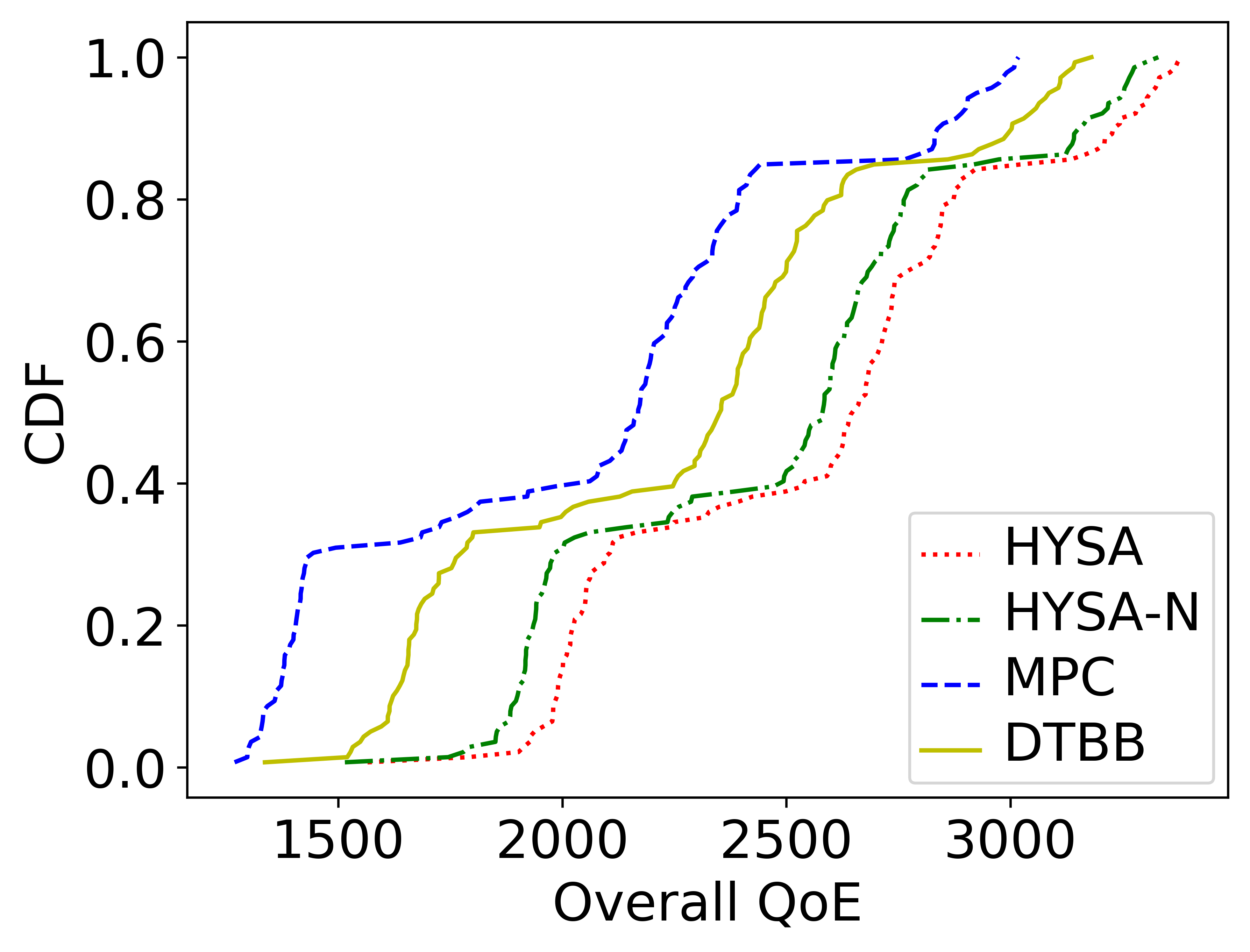}
    \end{minipage}
  }
  \subfigure[Room Video]{
    \begin{minipage}[b]{0.3\linewidth}
    \includegraphics[width=1\linewidth]{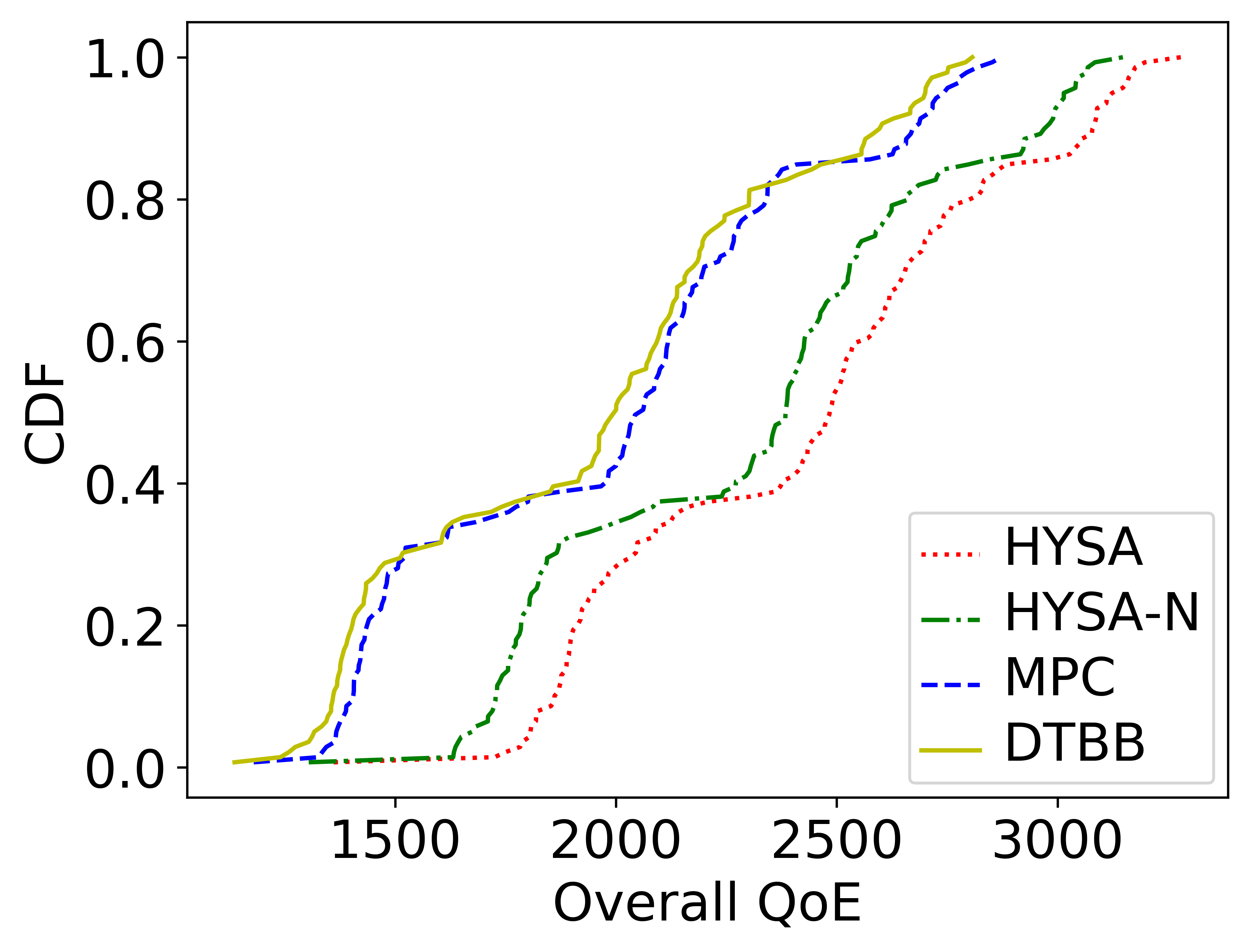}
    \end{minipage}
  }
  \subfigure[Game Video]{
    \begin{minipage}[b]{0.3\linewidth}
    \includegraphics[width=1\linewidth]{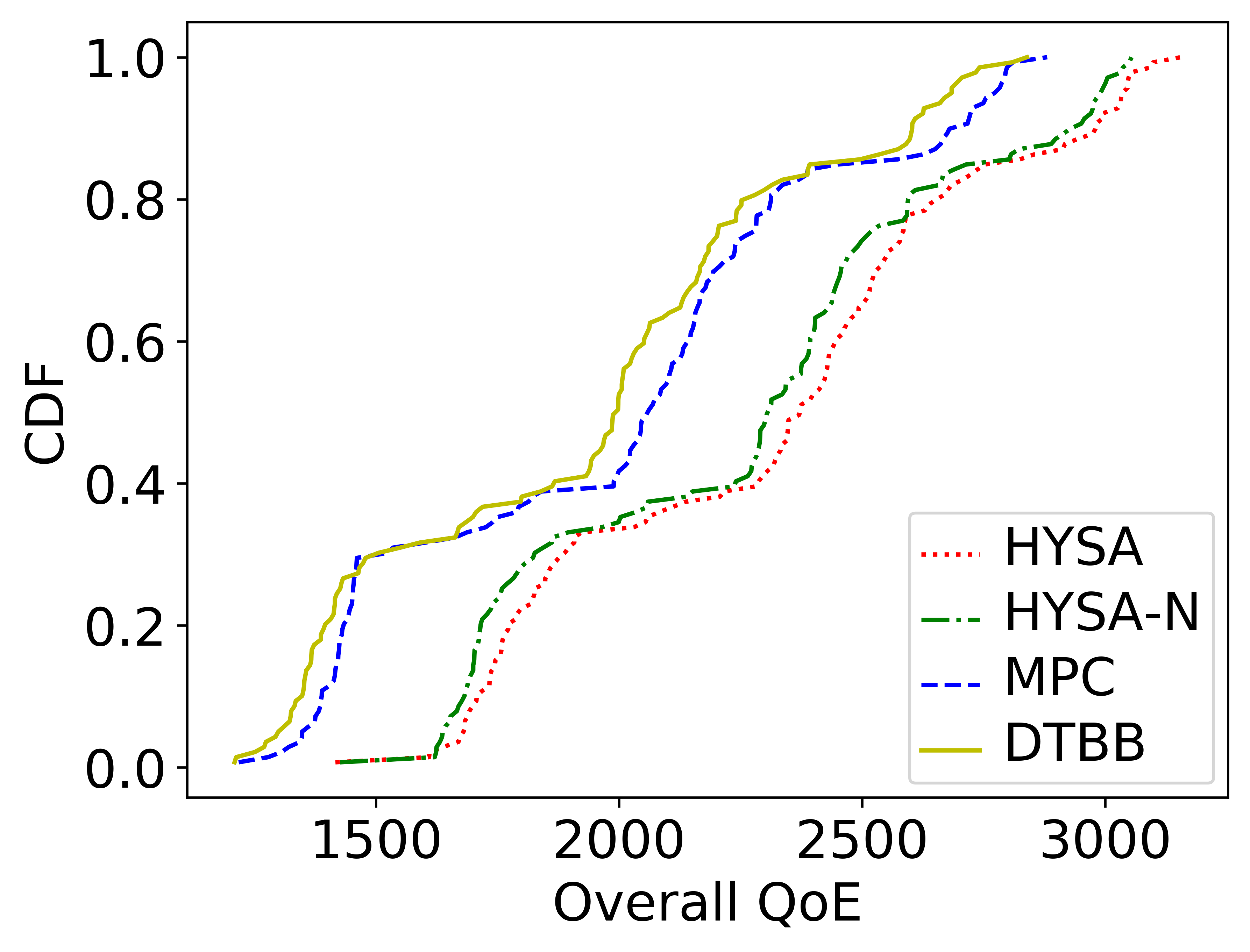}
    \end{minipage}
  }
  \caption{Overall QoE when streaming three types of videos}
  \label{qoe_of_three_videos}
  \Description{Simulation results when streaming three types of videos}
\end{figure*}

\subsection{Bitrate Control}
The latency-constrained bitrate control module makes bitrate decisions 
based on some state information, such as buffer occupancy, 
predicted segment bitrate and playback rate derived from the playback rate control module. 
The optimal bitrate is selected to minimize $D_{n+1}$, the estimated latency 
after downloading the next segment.

% We simply use Weighted Moving Average to estimate the future network throughput $\hat{C}_{n+1}$ 
% using $N_{2}$ historical throughputobservations.
% \begin{equation}
%   \hat{C}_{n+1} = \sum_{i=n-N_2}^{n} \alpha_{i} C_{i}
% \end{equation}
Denoting segment length by $d$,
the duration of downloading the upcoming segment n+1 at quality 
$m$, notated by $T_{n+1}$, can be calculated using the predicted segment bitrate $\hat{R}_{n+1,m}$ 
and the estimated network throughput $\hat{C}_{n+1}$ derived by Weighted Moving Average as follows:
\begin{equation}
  T_{n+1} = \frac{\hat{R}_{n+1,m} d} {\hat{C}_{n+1}}
\end{equation}

During the downloading process, the player will consume buffered video at 
playback rate $\gamma_{n+1}$ if the buffer isn’t drained. The buffer occupancy after 
downloading, namely $B_{n+1}$, can be calculated as:
\begin{equation}
  B_{n+1} = max[B_{n}+d-\gamma_{n+1} T_{n+1},0]
\end{equation}

To estimate the latency caused by video accumulated at CDN after downloading the next segment, 
we start with estimating the video accumulation speed at CDN, denoted by $\hat{v}_{n+1}$, as follows:
\begin{equation}
    \hat{v}_{n+1} = \beta v_n = \beta  \frac{N\_nst_{n}-N\_nst_{n-1}} {T_{n}} d_f
\end{equation}
where $\beta$ is a predictive factor, and $N\_nst_n$ indicates the index of the latest frame 
at CDN after downloading the $n$-th segment.  
Then, the latency caused by accumulated video at CDN after next 
downloading interval, namely $D\_cdn_{n+1}$, is calculated by:
\begin{equation}
    D\_cdn_{n+1} = max[(N\_nst_{n}-N\_dld_{n}) d_f + \hat{v}_{n+1}T_{n+1}-d,0]
\end{equation}
% \begin{gather}
% \begin{align}
%     D\_cdn_{n+1} = \ & max[(N\_nst_{n}-N\_dld_{n}) \tau_{frame}  \nonumber \\
%                     & + \hat{v}_{n+1}T_{n+1}-\tau_{segment},0]
% \end{align} 
% \end{gather}
where $N\_dld_{n}$ represents the index of most recently downloaded frame. 
The objective of the bitrate adaptive algorithm is to find the quality that 
results in lowest latency without interruptions, which can be described as 
the following optimization problem:
\begin{gather}
  \begin{align}
  Minimize \  D_{n+1} = & B_{n+1} + D\_cdn_{n+1} \nonumber\\
  Subject \ to \ & B_{n+1}>B_{th}
\end{align}
\end{gather}
where $B_{th}$ is a warning threshold indicating the upcoming stall event. 
Since there are finite available bitrates, we can get the optimal quality level $quality_{n+1}$ 
by going through all choices.

% 3.5 frame dropping control
\subsection{Frame Dropping Control}
The client can reduce the latency by dropping some frames when current latency is above a specific 
threshold. We propose a QoE-oriented frame dropping 
method to adaptively adjust the latency threshold that triggers frame skipping. Assuming that the client 
skips $N$ frames when the latency is above $l_{n+1}$ during next segment’s downloading, the positive and 
negative impact of frame skipping on QoE compared to non-skip, denoted by $QoE_{p}$ and $QoE_{n}$, 
can be estimated based on the given QoE model:
\begin{equation}
  QoE_{n}=p_{q} V_{n+1,quality_{n+1}} d_f N + p_s d_f N
\end{equation}
\begin{equation}
  QoE_{p}=p_{d} \lambda l_{n+1} N
\end{equation}
Here, $\lambda l_{n+1}$ is used to estimate the average latency if frame skipping 
is not performed when the latency is above $l_{n+1}$. 
Therefore, when $QoE_{p}$ is larger than $QoE_{n}$, frame skipping is a good choice:
\begin{equation}
  p_{d} \lambda l_{n+1} N > p_{q} V_{n+1,quality_{n+1}} d_f N +p_s d_f N
\end{equation}
\begin{equation}
  l_{n+1} > \frac{(p_{q} V_{n+1,quality_{n+1}} + p_s )d_f} {p_{d} \lambda}
\end{equation}

Thus, the latency threshold $latency\_limit_{n+1}$ that triggers frame skipping 
when downloading the next segment can be set as:
\begin{equation}
  latency\_limit_{n+1} = \frac{(p_{q} V_{n+1,quality_{n+1}} + p_s )d_f} {p_{d} \lambda}
\end{equation}

%%%%%%%%%%%%% Section4. Evaluate
\section{Evaluation}
In this section, we present the extensive evaluations, 
by which the following questions can be answered:
(1)	Is the segment bitrate prediction helpful?
(2)	How does HYSA compare to existing adaptation schemes?

To evaluate the performance of HYSA for streaming different types of videos, 
video traces in three scenes including room, game and sports are used for the evaluations. 
These videos are encoded at bitrates in $\lbrace 500, 850, 1200, 1850 \rbrace$ kbps.
Besides, our evaluations use 140 network traces sampled from real network scenario. 
The average bandwidth of these network traces covers from 0.8Mbps to 2.5Mbps, 
while variance covers from 0.1Mbps to 2.0Mbps.

First, experiments are conducted to evaluate the accuracy of KAMA-based segment 
bitrate prediction by calculating the prediction error $\frac{|PredictedBitrate-ActualBitrate|} {ActualBitrate}$. 
The results show that it can reduce the prediction error to 0.22, 
against the prediction error of 0.258 when using the segment's coding bitrate for prediction directly.

Then, we compare HYSA to the following adaptation schemes 
using the simulator mentioned previously: 
(1) HYSA-N: our baseline scheme where segment bitrate prediction is not included. 
% We employ it to evaluate whether the segment bitrate prediction contributes to making 
% better decisions or not. 
(2) MPC\cite{RobustMPC}: uses buffer occupancy and throughput predictions to select the bitrate which 
maximizes a given QoE metric over a horizon of five future chunks. 
(3) DTTB\cite{DTTB}: selects video bitrate based on a dynamic buffer threshold adapted according 
to the estimated throughput.
% \begin{itemize}
% \item {\verb|HYSA-N|}: our baseline scheme where segment bitrate prediction is not included. 
% % We employ it to evaluate whether the segment bitrate prediction contributes to making 
% % better decisions or not.
% \item {\verb|MPC|\cite{RobustMPC}}: uses buffer occupancy and throughput predictions to select the bitrate which 
% maximizes a given QoE metric over a horizon of 5 future chunks. 
% \item {\verb|DTTB|\cite{DTTB}}: selects video bitrate based on a dynamic buffer threshold adapted according 
% to the estimated throughput.
% \end{itemize}
Table \ref{QoE_metrics} provides the average value of overall QoE and other QoE metrics 
that each scheme achieves on the entire 
network traces and video traces. Figure \ref{qoe_of_three_videos} gives the 
Cumulative Distribution Function (CDF) of each network trace when streaming three types of videos.
From these evaluations, we can easily draw two conclusions. Firstly, the average overall QoE 
is improved with the KAMA-based segment bitrate predictions rather than the 
coding bitrates, which demonstrates that the KAMA-based segment bitrate prediction contributes to 
making better decisions. Secondly, the proposed hybrid control scheme outperforms other schemes 
with respect to overall QoE 
when streaming video in different scenes under various networks, because it tends to choose 
higher quality to improve bandwidth utilization, as illustrated by Table \ref{QoE_metrics}.

\section{Conclusion}
In this paper, we have presented HYSA -- an effective hybrid control scheme consisting of 
heuristic playback rate control, latency-constrained bitrate control and QoE-oriented 
adaptive frame dropping. 
Our algorithm adopts Kaufman’s Adaptive Moving Average to predict the segment bitrates, 
with which we could make the bitrate decisions more accurately.
Extensive simulation results have demonstrated that the segment bitrate 
prediction is advantageous in making better 
decisions, and HYSA can achieve higher overall QoE than the prior state-of-the-arts.

%
% The acknowledgments section is defined using the "acks" environment (and NOT an unnumbered section). This ensures
% the proper identification of the section in the article metadata, and the consistent spelling of the heading.
\begin{acks}
This work was supported in part by the National Natural Science Foundation of China (Grant No. 61971382)
\end{acks}

%
% The next two lines define the bibliography style to be used, and the bibliography file.
\bibliographystyle{ACM-Reference-Format}
\bibliography{reference}

\end{document}